# How Fluffy is the Cloud?: Cloud Intelligence for a Not-for-profit Organisation

**Purva Koparkar**
Faculty of Business Government & Law
University of Canberra
Canberra, Australia
Email: purva.koparkar@canberra.edu.au

**Dale MacKrell**
Faculty of Business Government & Law
University of Canberra
Canberra, Australia
Email: dale.mackrell@canberra.edu.au

## Abstract

Business Intelligence (BI) is becoming more accessible and less expensive with fewer risks through various deployment options available in the Cloud. Cloud computing facilitates the acquisition of custom solutions for not-for-profit (NFP) organisations at affordable and scalable costs on a flexible pay-as-you-go basis. In this paper, we explore the key technical and organisational aspects of BI in the Cloud (Cloud Intelligence) deployment in an Australian NFP whose BI maturity is rising although still low. This organisation aspires to Cloud Intelligence for improved managerial decision making yet the issues surrounding the adoption of Cloud Intelligence are complex, especially where corporate and Cloud governance is concerned. From the findings of the case study, a conceptual framework has been developed and presented which offers a view of how governance could be deployed so that NFPs gain maximum leverage through their adoption of the Cloud.

## 1 Introduction

Cloud computing is a relatively recent term initiated by industry practitioners. It is a facility which promotes on-demand, ubiquitous, and secure networked access to a shared pool of configurable computing resources for rapid provisioning and release, with minimal managerial effort and service provider interaction (Mell and Grance 2009). The Cloud is becoming an attractive option for small to medium-sized enterprises (SMEs) and small not-for-profit (NFP) organisations for whom even basic information technology (IT) remains an obstacle. For these firms, hardware and software infrastructure is not sufficiently funded and often out-dated with a heavy reliance on non-IT staff for IT operations (Lacity and Reynolds 2014; Strickland et al. 2010). Business Intelligence (BI) in the Cloud, as Cloud Intelligence, integrates BI systems with Cloud computing architecture as an adaptable and affordable computing platform to improve the quality and speed of business decisions (Baars and Kemper 2010).

This paper examines deployment options for migration to the Cloud as well as security, governance and risk management aspects for consideration before a Cloud Intelligence solution is selected and adopted. Cloud-oriented concepts are operationalised in a case study of a small Canberra-based NFP organisation which is steadfastly moving its data, technical and business processes to the Cloud. The case study aids in comprehending the cautious but determined steps of a NFP in implementing Cloud Intelligence along with the outcomes, both expected and unexpected.

Findings from the paper are both practical and theoretical. They demonstrate how a NFP organisation relies on the Cloud provider to shoulder governance responsibilities when the significance of developing Cloud governance policies in-house is not well appreciated. A conceptual framework is presented which suggests a possible option for how governance responsibilities could be shared with the Cloud provider allowing the NFP to gain maximum leverage from their adoption of the Cloud.

## 2 LITERATURE REVIEW

For the literature review, we draw from scholarly and industry literature which together form a holistic background of the research domain and practice setting in which the study took place.

## 2.1　Overview

Business Intelligence in the Cloud (Cloud Intelligence) represents the coming together of two key IT trends: Cloud computing architecture as a flexible and cost effective computing platform and BI technology as a support for swift organisational decision making (Goel 2010). Cloud Intelligence promises the delivery of business and consumer services over the Internet in a responsive, scalable and economical manner. The Cloud computing environment enables BI to be distributed as three main delivery service models: Software as a Service (**SaaS**) whereby software applications run in the Cloud; Platform as a Service (**PaaS**) which provides shared development and run-time platforms and manages underlying software and hardware layers; and Infrastructure as a Service (**IaaS**) such that storage and computing power are offered on-demand as a service (Lacity and Reynolds 2014; Marston et al. 2011). The virtual environment provides users with the capability to access computing power which they may not have been able to access earlier due to financial or organisational limitations (Tamer et al. 2013). Notwithstanding the hype, there are impediments to implementation. These are discussed in this paper along with the merits of Cloud Intelligence.

## 2.2　Cloud Intelligence Deployment

Descriptions of Cloud deployment models are readily available. Marston et al. (2011, p.180) provide an overview, explaining that

> *… **public** cloud such as Google Apps is characterised as being available from a third party service provider via the Internet, and is a cost-effective way to deploy IT solutions, especially for SMEs. A **private** cloud offers many of the benefits of a public cloud computing environment, such as being elastic and service based, but is managed within an organisation. Private clouds provide greater control over the cloud infrastructure, and are often suitable for larger installations. Finally, a **hybrid** cloud is a combination of a public and private cloud — typically, non-critical information is outsourced to the public cloud, while business-critical services and data are kept within the control of the organisation.*

The strategy for migration of BI to the Cloud begins by analysing and evaluating the present business and designing the Cloud Intelligence solution accordingly (Baars and Kemper 2010). Cloud deployment models provided by various Cloud Intelligence providers should be evaluated and selected based on factors like performance requirements and existing interdependencies, network costs, security and privacy essentials. Implementation of the solution toward the Cloud may be achieved in iterative stages, through continuous transmission of data, services and processes (Juan-Verjedo and Baars 2013).

In a BI environment, there are several ways that organisations can mix and match on-premises with the Cloud to enhance existing architecture. A common scenario is that a SaaS or PaaS BI solution is purchased from a Cloud Intelligence vendor. The organisation's source systems and data warehouse remain in the corporate data centre but the BI solution and associated data mart run in the Cloud. Organisations are required to pay the costs of creating a custom data mart and also to build a relatively complex, custom BI environment. Also, they still need to transfer data to the Cloud and users may experience response time delays due to Internet latencies (Juan-Verdejo and Baars 2013).

Another scenario is to put the entire data warehousing environment in the Cloud as an IaaS, which makes sense if operational applications also run in the Cloud. This scenario may apply only to a few companies that have fully embraced Cloud computing for all application processing. One of the concerns of this scenario could be regarding vendor lock-in. This architecture could prove dominant in the future once Cloud Intelligence gets past security and latency hurdles (Tamer et al. 2013).

## 2.3　Security in the Cloud

Currently many organisations consider security and privacy as major concerns when it comes to using the Cloud. There is a tendency to assume data is safer in a corporate data centre or in-house where the organisation has more control. In a Google (2012) white paper, it is claimed that Cloud computing can be as secure, if not more secure, than traditional environments because most businesses do not have the security intelligence gathering capabilities that are available to Cloud providers with resources for significant security measures. Hence, for SMEs and NFPs, security in the Cloud could be a huge benefit as they are able to adopt enterprise-level security which has been denied to them earlier through prohibitive costs.

The view presented in the Executive Report prepared by IT Business Edge for Hosting.com (2010) is that the vast majority of potential Cloud vulnerabilities such as phishing, downtime, data loss, password weaknesses, and compromised hosts running botnets are similar, if not identical, to those found in IT architecture across the board. This argument is supported by Chen et al. (2010) in which they remark that many of the concerns described as 'Cloud security', in fact, simply reflect customary Web-application and data-hosting problems because the underlying issues remain well-established.

The Cloud provider is obliged to address client concerns. These include the following: the client organisation should have a clear understanding of how data is secured at rest on storage devices; how data is secured during transit; what mechanisms are used to authenticate users; how legal and regulatory issues are addressed; issues of multi-tenancy in the Cloud and the separation of data and applications between customers; and issues of backup and recovery from incidents (Hanna and Molina 2010).

## 2.4   Cloud Governance

IT Governance is required for putting structure around how organisations align IT with business strategy, to ensure that companies are on track for achieving long term goals, and for implementing efficient ways to measure IT performance (Prasad et al. 2013). The major focus areas for IT governance are strategic alignment, value delivery, resource management, risk management and performance management (ITGI 2009). A governance model provides structure for managing multiple projects, avoiding redundant effort, and allowing organisations to exploit investments in services by managing the service portfolio (Stahl et al. 2010).

The Cloud environment is very different from traditional outsourcing and requires an extension to existing IT governance frameworks to include the unique elements of applying IT governance to third party service providers as well as security and compliance requirements across physical, virtual and Cloud environments through appropriate Service Level Agreements (SLAs). Cloud governance must satisfy legal and industry standards, a concern recognised by many organisations because control regarding a number of issues must be relinquished to the Cloud provider in order to access the Cloud infrastructure (Farrell 2010). During the adoption of Cloud computing, despite the handover of certain IT functions, one school of thought is that responsibility around governance should remain in-house since governance is an operating model for how organisations make decisions about the acquisition and use of IT based on their overall IS strategic plan (ISACA 2013).

Existing IT governance frameworks like COBIT, ITIL, ENISA (IT security), COSO-ERM (risk) and international standards like ISO 38500 (IT governance) and ISO 27000 series (IT security) can be useful for forming Cloud governance policies and controls. In the case of organisations which have an existing IT governance framework, the amount of effort necessary to extend this framework to the Cloud will be less because core IT management disciplines have not changed, just shifted, from the IT organisation to the Cloud service provider (Bailey and Becker, 2014).

Without effective Cloud governance, a Cloud initiative can become shadow IT. This is an even greater issue since a Cloud solution may not be in line with an organisation's requirements for control, documentation, security, and reliability. This also creates a high risk of getting locked in to a Cloud vendor, or the uncontrolled adoption of different Cloud services might lead to costly redundancies and incompatibilities (Rebollo et al. 2012). Hence, the key to effective selection and implementation of Cloud services may be a robust Cloud governance structure which is in line with the overall corporate governance structure of the organisation.

These and other issues in the literature review are identified in the case study of the local NFP discussed below.

## 3   Not-for-profit Case Study

The case study is Connections ACT Incorporated, a local Canberra community NFP organisation operating in the Australian Capital Territory (ACT) committed to helping disadvantaged communities and families who are homeless (or at risk of becoming homeless). Connections ACT consists of 12 paid staff that assist over 3,000 people in the ACT annually through two services: First Point, the central intake service for the homeless, and CanFaCS, a service for homeless fathers and their children. The researchers have a trusting and continuing association with Connections ACT and have witnessed their transitioning towards Cloud Intelligence.

Like many small NFPs, the managers at Connections ACT struggle with onerous compliance and ad hoc reporting since client data capture is often cumbersome, data quality is poor, software is inadequate, and silos of data exist in the organisation due to non-linking databases and a reliance on spreadsheets, in turn thwarting the extraction and use of the valuable information they contain. NFPs such as Connections ACT are increasingly reliant on accurate and quickly retrievable organisational data for regulatory reporting and on-going funding. This is difficult to realise for most NFPs since the reporting software available is predominantly suited to organisations in the for-profit sector which have different needs and resource-bases, and are not liable to compliance reporting to the same degree as NFPs.

In 2012, when deprived of physical access to offices and computers due to a localised fire emergency, managers at Connections ACT recognised that it was vital to deploy the Cloud for virtual hosting of their databases, software applications, and e-mail. Thus began the process of migration to BitCloud. From the requirements elicitation phase of the project in 2012, it was determined that a practical outcome of the project would be the design, development and deployment of stable and reliable data repositories, being a preliminary phase to BI. The migration of non-structured data from spreadsheets and a whiteboard to a SQL database showed the growing recognition of the importance of data residing in a structured format. Indicative of an emerging BI maturity was the design of a data mart to integrate data from various sources and systems. The case study is now an ongoing longitudinal project with further details available in MacKrell and McDonald (2014).

During the requirements elicitation phase in 2012, nine face-to-face, in-depth semi-structured interviews were conducted with senior Connections ACT managers and one Board member, as well as three group meetings with University of Canberra BI and data warehouse experts to ascertain the status of technology and information needs in the NFP. From when the project began in 2013, meetings with Connections ACT managers and users as the client, and interviews with ICT student teams as software developers have been conducted at several points in the semester: at the start and end of each semester, mid-semester and/or any crises point. These interviews and meetings are face-to-face, semi-structured, conversational, and conducted with a single person or the whole student team. While interviews and meetings are comparatively informal, they are guided by a prepared interview script or meeting agenda in order to ensure all topics are covered. Most interviews and meetings have been recorded then transcribed, resulting in about 50 hours of interview data to build a comprehensive and longitudinal account of the project.

Organisational and Board documents have provided a complementary triangulated perspective of the transition in Connections ACT towards Cloud Intelligence. For this paper, an additional meeting, supported by Board papers, was conducted by both researchers with the Executive Officer (EO) and the Business Manager of Connections ACT. Excerpts from this particular meeting have been provided to illustrate pertinent points.

## 4    Analysis and Discussion of Findings

Cloud computing facilitates the acquisition of custom solutions for NFPs at affordable and scalable costs on an elastic, pay-as-you-go basis (Strickland et al. 2010). Furthermore, if NFPs like Connections ACT are willing to engage the services of university ICT students in work-integrated learning (WIL) projects, then this is a win-win-win situation (Strickland 2008). It means that NFP organisations can acquire Cloud-based IS at low cost, the students gain practical workplace experience, and educators and academics stay more relevant by being involved in real-life work situations.

As Strickland et al. (2010) note, the Cloud has similar relevance for most SMEs and small NFPs due to critical IS that span several areas of IT use: applications for client management, data storage, e-mail hosting, as well as office productivity and collaboration tools. Assuming the differences in Cloud usage between SMEs and small NFPs are minimal, then Connections ACT can be considered as representative of both. Cloud-based computing is the means by which many of these under-resourced, financially struggling organisations can advance towards sustainable hardware infrastructure and software tools for monitoring the delivery of services and programs, and to support the preparation of compliance and standard reporting. In this section, we look at how Cloud security, governance and risk are perceived by Connections ACT management along with the benefits and the challenges in the shift towards a complete Cloud Intelligence solution.

### 4.1    Selecting the Cloud Provider

Before migrating to the Cloud, Connections ACT had a history of individual desktops that were not linked together. Connections ACT was in need of better computing systems due to a series of IT disruptions and the fact that existing server infrastructure was inadequate and in need of replacement. The options considered were to either move Connections ACT data and applications into a full Cloud-based repository with backup provided or to run applications from the Cloud but have locally based backup.

Any Cloud solution should be chosen only after weighing up the various arguments before embarking on the Cloud. It is imperative that organisations be diligent in mitigating the risks by operating on a trusted computing platform and being selective concerning the types of data chosen to store in the Cloud (Tamer et al. 2013). Connections ACT management were cognisant of regulatory compliance requirements and only those Cloud providers who had secondary backups within Australia were shortlisted. Since the client data stored is very sensitive, a high level of security is crucial, which led to the decision to use the private Cloud with dedicated IP and additional data security for all data storage and applications. Not having to maintain and manage hardware on-site and the scalability feature of the Cloud led to the decision of migrating fully to the Cloud. After comparing various aspects like initial and ongoing costs, technical support, backup and disaster recovery options, physical, network and transmission security, the managers from Connections ACT selected BitCloud as their trusted provider.

Connections ACT management had done their homework well and two testimonies on BitCloud services were taken into account while arriving at this decision. As the EO explained:

> *We chose BitCloud specifically because they were the only Australian company at that time that had off-site backups in Australia and complied with Australian privacy legislation and they were cost effective and had credibility. We'd done due process, and due diligence checks on references.*

### 4.2 Deployment and Security

The deployment option chosen by the managers at Connections ACT was to fully embrace the Cloud, locating data and applications in the private Cloud. Concerns about vendor lock-in which are typically associated with this deployment option were addressed by having a one monthly renewable contract with the Cloud provider. This left Connections ACT management with the option of changing providers if and when their requirements change. For the moment, managers at Connections ACT are very satisfied since BitCloud delivers in a timely manner precisely what is expected. The EO appreciates the fact that the Relationships Manager at BitCloud, a NFP expert, can talk them through and help in decision making at times when Connections ACT staff members are not exactly sure what they want or need.

One of the concerns with the earlier pre-2012 in-house arrangement was that IT security was poor and virus protection was weak. This concern was eliminated by BitCloud taking responsibility for security of the sensitive client data as well as software applications. All data is accessed through a private network by using a secure VPN.

> **EO:** *People get their own log-on and user accounts and go through the VPN network and one can't access anything - VM1 or VM2 - except through the VPN.*

Even the virtual development and test environment on VM2 used by students can be accessed only through secure VPN connection. At no time, can students access the VM1 and risk interfering with current operations and data.

### 4.3 Risk and Governance

In the case of Connections ACT, BitCloud assumed accountability for governance through the SLA. While this appears to leave Connections ACT somewhat exposed, it actually frees the NFP to pursue core business. At this point in the interview, for confirmation and reassurance, the EO examined the SLA.

> **Researcher:** *Do you comply with ISO 27000 or any other security standard?*
>
> **EO:** *BitCloud probably does [comply] but we don't have to because they're [BitCloud] doing it for us.*

As part of the arrangement with BitCloud, all data is backed up once every 24 hours and mirrored into two data storage warehouses, one in Melbourne, one in Sydney. Risk assessment has been done that one day's work will be lost if there is any data centre outage or failure. Having two backups at different locations satisfies the disaster recovery needs of Connections ACT at a comparatively low cost compared to on-site backup. In the event of a state-wide disaster, Connections ACT is reliant on support from the government. As the EO explained:

> *But if there's an ACT-wide black out, then we've spoken to the directorate [ACT Government] and they've said, we've got a business continuity plan, we can guarantee that we will give*

> *you premises and electricity in the event of a disaster, but we can't share that business plan with you because it's government in-confidence.*

Migration of all IS to BitCloud has eliminated the risk of being physically locked out of their premises. Management at Connections ACT understand that it is well and good to be on the Cloud, but if the office does not have an Internet connection, it is a big risk mitigated by using a reliable Internet provider.

### 4.4  Meeting Expectations

So it is that Connections ACT have adopted Cloud delivery services of Iaas, Saas and PaaS using a private Cloud deployment model. The arrangement with BitCloud has fulfilled all expectations, like having access to the latest technology, systems and applications without having to invest in expensive hardware, software or licenses, and to retain skilled IT staff for maintenance and regular upgrades to infrastructure. Connections ACT now have predictable expenses through monthly subscriptions which are much lower than their earlier expenses, calculated to be less than half the cost of hiring a full time staff member. Another benefit is the high level of customer expertise and support that the staff receive and the ease with which IT needs are met. The EO clarified this by saying:

> *We've got a direct line. We just ring him [BitCloud rep]. He's in Sydney. He talks to us. It's almost like having someone on staff in a way.*

Remote access has been very helpful as the Business Manager from Connection ACT indicated:

> *With a flexible workplace you've got the option now of staff being able to work from home if they need to, if they've got caring responsibilities and the team take out iPads and they can sit there and update SHIP [client database] while they're at an appointment, search for information. [It] makes their job a lot easier.*

Features like 24/7 access to organisational data and software tools are immensely useful when a community organisation has previously faced a debarment from their premises and computers through no fault of their own, as Connections ACT had experienced. BI artefacts like reliable data repositories and management systems for both First Point and CanFaCS being all in one place are a bonus. These steps towards a BI solution have been helpful in fulfilling regulatory compliance requirements.

In summary, since the decision to deploy BitCloud for migrating to the Cloud, staff at Connections ACT have received high quality service and support from the provider. BitCloud has assumed all responsibility for IT infrastructure requirements of Connections ACT as well as its security and governance at a reasonable monthly charge. The Cloud allows flexibility for changing requirements and also provides the opportunity to experiment with the latest software applications through subscriptions. One other important aspect is remote access which addresses some of the business continuity and risk management planning activities for Connections ACT.

## 5  Concerns Illustrated with Governance Frameworks

Overall the Cloud Intelligence solution is a constructive one for Connections ACT however there is little room for complaisance. The statement below by the EO at Connections ACT clearly indicates that management has adopted a 'set and forget' approach regarding their migration to the Cloud.

> *Well, your questions around governance really indicate that we've just done a 'set and forget'. We initially identified what the issues were and what the advantages were for going to the Cloud. We did our research at the time, but we haven't done a comparative analysis of that recently. We haven't reviewed ... should we be using a different provider, changing the existing arrangement or any sorts of governance arrangements.*

Bearing in mind the lack of dedicated resources in many SMEs and NFPs to satisfy IS needs, this could be a common trend where organisations transfer, with some relief, responsibility to the Cloud provider in terms of infrastructure, access, support and service, security as well as governance. With regards to risk, often a basic pros and cons assessment is done before migration to the Cloud while an ongoing assessment of the selected Cloud solution may not be performed. Figure 1 is our representation of how many SME/NFP organisations tend to view the responsibility shift to the Cloud provider when they adopt the Cloud.

While this structure could be perceived as being helpful in freeing the managers and staff of Connections ACT from IT responsibilities and providing them with more resources to focus on their core business of providing shelter for the homeless, it raises concerns regarding alignment of the selected Cloud solution with strategic business needs. Uncritically trusting the Cloud Provider to make all IT-related decisions may lead to a loss of internal IT knowledge. This was noted in a statement by the Business Manager at Connections ACT:

> *There are so many times when we don't know what we want and they [BitCloud] talk us through it and help us decide.*

While high levels of trust can be perceived as positive, on the other hand, loss of IT knowledge hinders an organisation from realising the full potential of Cloud services and also locks in the organisation as they fear they may not get similar support elsewhere. Furthermore, while adoption of the Cloud eliminates the need to have extensive in-house technical/operational IT knowledge, comprehending the various functionalities of the Cloud from a business perspective is often crucial for continuing Cloud success. Also lack of governance leads to having unclear indications of where the organisation is heading in terms of strategic goals like BI in the Cloud. This might possibly be one of the reasons why some SMEs or NFPs continue on their journey towards a BI solution but fail to reach their supreme objective of a successful implementation.

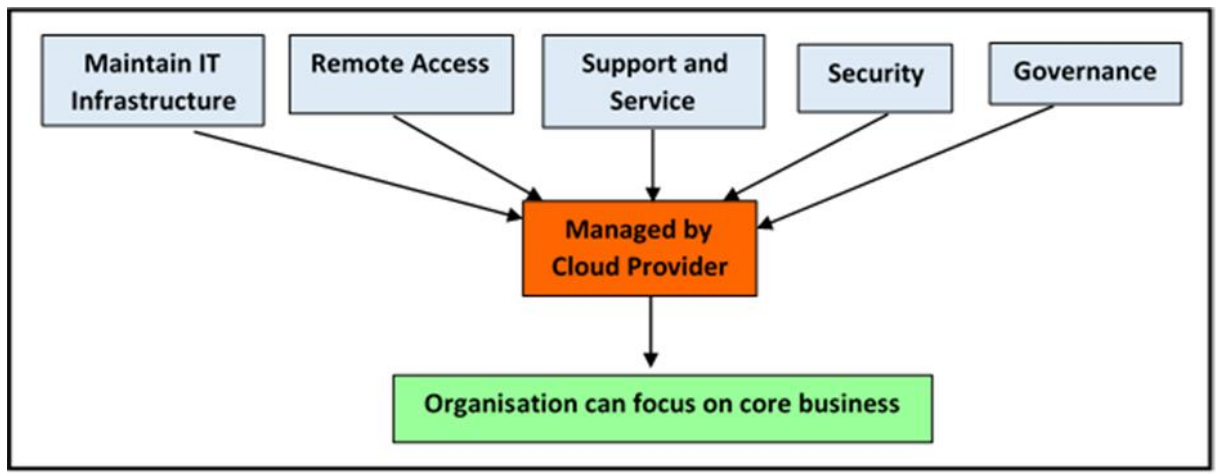

*Figure 1: AS-IS Structure for SMEs and NFPs*

The main concerns of most organisations, namely security, risk, and business/IT alignment, can be addressed through governance. A strong governance strategy and management plan is the recommended solution to gain maximum benefit from Cloud initiatives (Crowe Horwath al. 2012). This should set the objectives and direction for Cloud computing through aligning IT with the goals of a BI solution and thus adding value to the organisation. Many companies may have some sort of corporate governance in place, but very few understand the need for Cloud governance. By understanding the basics of Cloud governance, the organisation would be in a better position to exploit investments in services by managing the service portfolio and identifying potential investment areas, in addition to measuring how they are tracking along their business needs and long term strategic goal of BI (Prasad et al. 2013).

For a SME or NFP, adopting governance frameworks like ITIL or COBIT may not be suitable due to the small size of the organisation as well as the lack of available resources, but for effectively implementing and managing Cloud services, the organisation needs to have well-defined policies regarding information security management, risk and compliance management, and application lifecycle management. Some controls from international standards like ISO 38500 or the ISO 27000 series may be suitable for SMEs or small NFPs as a starting point to form Cloud governance policies. At the least, the organisation needs to have an unequivocal understanding of the 'what' aspect from their end and the Cloud provider could primarily be responsible for the 'how' aspect.

Bailey and Becker (2014) advocate that Cloud governance requires defining policies that help to outline the responsibilities for IT management, business processes, and applications. In Figure 2, we propose a conceptual framework of a way in which responsibilities can be divided so that a SME or small NFP organisation can gain maximum leverage from migrating to the Cloud. The proposal is that the Cloud provider manages everything to ensure smooth functioning of IS which includes, but is not limited to, the infrastructure, access, support service as well as instigating the security controls at the physical and logical level. The organisation, on the other hand, assumes responsibility for developing an unambiguous Cloud governance policy focused on risk, security and IT alignment. The level of Cloud governance would vary according to organisational size, industry or applicable regulations.

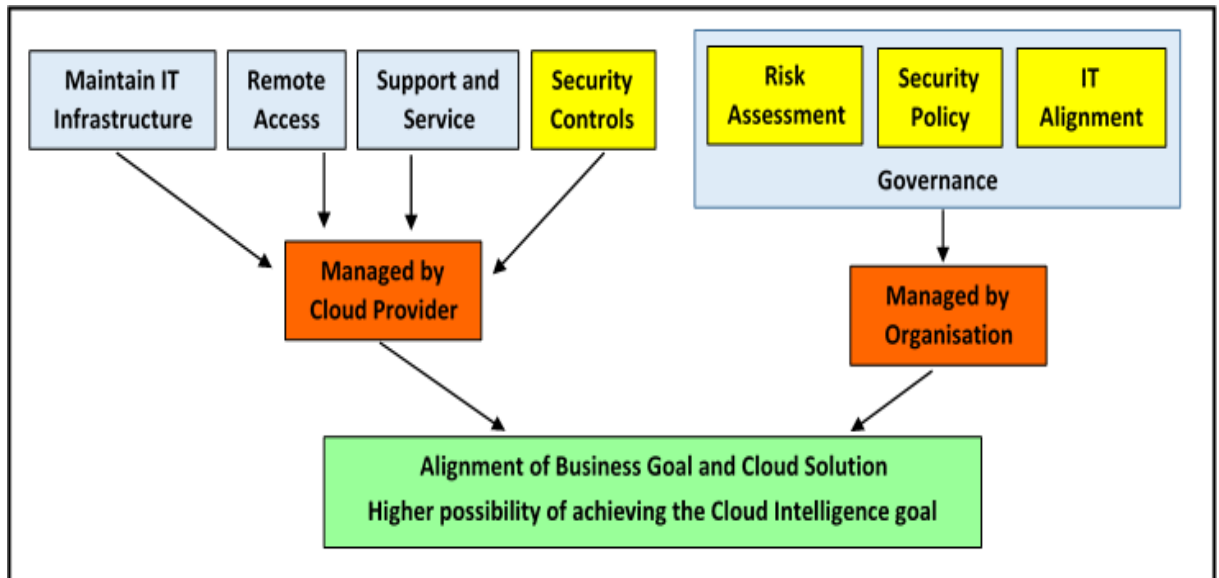

*Figure 2: Alternative Structure for SMEs and NFPs*

## 6　Conclusion

Cloud computing offers the opportunity for SMEs and small NFPs to re-examine their approaches as to how BI may be used to support their missions (Strickland et al. 2010). Cloud computing is rapidly emerging as an attractive IT option for businesses with deficiency of knowledgeable IT staff and ageing computing infrastructure. This is due to many reasons, not least the cost advantages it offers to access IT resources remotely and to meet the fluctuating demand for those resources. While security in the Cloud remains a top concern for many businesses, the scope of security in Cloud computing is indistinct as it can be either an advantage or drawback to many organisations, depending on the sensitivity of the data and the criticality of the functions (Marston et al. 2011).

The main benefits of using BI in the Cloud are lower capital expenditure in infrastructure and lower upfront implementation costs along with the ability to deploy a recognised standard solution. That aside, it is important for organisations to understand that Cloud Intelligence implementations require custom data design and development which involves integrating data from multiple operational sources. A Cloud Intelligence solution may be a feasible answer to the challenges of the economic crisis, due to its flexible cost structure and scalability, but it comes with risks and vulnerabilities related to organisation's data, personnel, and reputation. Due diligence should be conducted on entities with whom the organisation will be engaged and using a trusted computing platform is obligatory.

In this paper, we argue that Cloud governance is essential in this process to align organisational goals with the benefits that can be gained from Cloud deployment. We present a conceptual framework in Figure 2 which suggests that responsibility of governance should remain within the organisation, although the responsibility of managing other IT functions may usefully be handed over to the Cloud provider. Cloud governance facilitates a better fit of Cloud computing services into existing processes of organisations to achieve business and financial objectives. Cloud governance assists to maintain a centralised decision-making process which is in-line with the overall strategy of the organisation. Governance is not something that can be considered as 'nice-to-have', it is something that every organisation 'needs-to-have' (Rebollo et al. 2012). There are undoubtedly a number of risks and uncertainties in transitioning to the Cloud, so strong governance and control are an essential part of any decision to move to the Cloud.

This paper represents in one document some of the issues along with governance, security, and risk management issues associated with Cloud computing before making any decision about implementing Cloud Intelligence. A major contribution of this study would be towards the practical aspects of understanding the technological needs of a NFP organisation and tackling accordingly the security concerns and threats that arise from using the Cloud. As more and more data moves from on-premises to the Cloud, it will become more feasible for NFPs and SMEs to deploy BI in the Cloud.

One of the limitations of this paper is that the frameworks, Figure 1 (drawn from the literature) and Figure 2 (adapting Figure 1 after analysing the interview data), have not been validated against supporting data at an operational level. Further research would be required to provide evidence of the effectiveness and feasibility of having in-house Cloud governance by SMEs and small NFPs. Nevertheless, the views presented in this paper about the responsibilities of governance have the potential to stimulate debate. The questions for future research could be: "how does assigning responsibility of governance to the Cloud provider allow an organisation to focus on their core business ?" or "how does this cause interference with the alignment of the Cloud solution to corporate strategic goals ?".